\documentclass[%
 reprint,
superscriptaddress,
 amsmath,amssymb,
 aps,
pra,
]{revtex4-2}

\usepackage{graphicx}
\usepackage{dcolumn}
\usepackage{bm}
\usepackage{hyperref}
\usepackage[mathlines]{lineno}
\usepackage[braket]{qcircuit}
\usepackage{pgfplots}
\usepackage{tikz}
\usetikzlibrary{decorations.text}

\DeclareMathOperator*{\argmax}{arg\,max}

\newcommand{\figref}[1]{Fig.~\ref{#1}}
\newcommand{\secref}[1]{Sec.~\ref{#1}}

\usepackage[bottom]{footmisc}

\begin{document}

\preprint{APS/123-QED}

\title{Improving quantum dot based single-photon source with continuous measurements}

\author{Anirudh Lanka}
\email{alanka@usc.edu}
\affiliation{Ming Hsieh Department of Electrical \& Computer Engineering, University of Southern California, Los Angeles, California}
\author{Todd Brun}
\email{tbrun@usc.edu}
\affiliation{Ming Hsieh Department of Electrical \& Computer Engineering, University of Southern California, Los Angeles, California}
\affiliation{Department of Physics \& Astronomy, University of Southern California, Los Angeles, California}
\affiliation{Department of Computer Science, University of Southern California, Los Angeles, California}

\date{\today}

\begin{abstract}
We propose a technique to improve the probability of single-photon emission with an electrically pumped quantum dot in an optical microcavity, by continuously monitoring the dot’s energy state and using feedback to control when to stop pumping. The goal is to boost the probability of single-photon emission while bounding the probability of two or more photons. We model the system by a stochastic master equation that includes post-measurement operations. Ideally, feedback should be based on the entire continuous measurement record, but in practice, it may be difficult to do such processing in real-time. We show that even a simple threshold-based feedback scheme using measurements at a single time can improve performance over deterministic (open-loop) pumping. This technique is particularly useful for strong dot-cavity coupling with lower rates of pumping, as can be the case for electrical pumping. It is also numerically tractable since we can perform ensemble averaging with a single master equation rather than averaging over a large number of quantum trajectories.
\end{abstract}

\maketitle

\section{Introduction}

Single-photon sources that produce high-quality photons on demand would be a technological development of great importance \cite{Aharonovich2016SolidstateSE, PhysRevLett.89.187901, arakawa2020progress}. Such sources have a myriad of applications spanning quantum information processing and computing, efficient quantum key distribution (QKD) \cite{Kupko_2020, Wang_2019, Lo_2014}, metrology, and quantum imaging, among others. In this work, we are particularly interested in linear optical quantum computation (LOQC) for which a key requirement is the availability of high-quality single photons.  Quantum processors utilizing photonic integrated circuitry (PIC) could be operated at room temperature and would not require expensive cryogenic cooling due to their weak coupling to the external environment \cite{universal_pqc}. In addition, photons propagate at the speed of light and offer a large bandwidth for data transmission, making them efficient data carriers. Hence, realizing a perfect single-photon source with ideal emission characteristics is very much desired. While there exist many methods to produce single photons, a quantum-dot-based approach is of particular interest due to their high excitation and collection efficiency\cite{uppu2020chip}.

For useful quantum information processing, an ideal photonic quantum computer should have single photon sources that satisfy three main properties. First, the probability to generate a single photon should be close to one, while keeping the multi-photon emission rate below a maximum tolerable level (two goals that may be somewhat in conflict). Second, the single photons emitted should be indistinguishable, which means that dephasing should be suppressed \cite{HighPerformanceQDot, IndistinguishablePF, InterferenceAC, PhysRevApplied.13.034007}. Indistinguishability requires that each photon emitted be absolutely identical to every other: they must have the same polarization, spatial mode, and temporal profile. This property is particularly crucial for applications such as boson sampling \cite{complex_optics}. A method to improve indistinguishability using continuous weak measurements was presented in \cite{PhysRevA.79.033831}. Third, the emitted photons must be collected with near-unit efficiency in the preferred quantum channel. There are various processes that affect this: pumping the quantum dot, emission into the cavity, cavity leakage to non-waveguide modes, and photon loss, among others\cite{sps&d, sps}. Although most of these processes depend on fabrication techniques and the quality of materials used, the duration of pumping can be controlled and optimized, which is the main focus here.

Photons are generated from a quantum dot by pumping it (either optically or electrically) under favorable bias conditions for a finite duration. The number of photons released depends on the duration of the pumping. Optical pumping is more straightforward experimentally; electrical pumping is better suited for large-scale integration. Also, as electrical pumping does not directly insert photons into the cavity, it opens up the possibility of pumping directly into an energy level resonant with the cavity mode, which can reduce timing uncertainties and improve purity in photon emission \cite{PhysRevA.79.033831}. In electrical pumping, a bias voltage is applied on a $p$-$i$-$n$ diode in which the insulator region contains a quantum dot embedded in an optical microcavity. This allows for an electron to tunnel through from the $n$-region to the dot. Further pumping will then enable a hole to tunnel through from the $p$-region to the dot and recombine with the electron. Ideally, this is when the pumping should be stopped so that no new electrons or holes tunnel. After a successful recombination event occurs, the dot will release a photon into the microcavity where it ultimately leaks into an external waveguide.

In the case of strong pumping---as optical pumping often is \cite{opt_v_elec_pump}---the duration of pumping can be short compared to the emission time of the dot. This simplifies the pumping process, but to avoid scattering extra photons into the microcavity one must generally pump the dot to a nonresonant state and rely on incoherent processes to drop into the resonant state \cite{opt_pump}. These incoherent processes increase timing uncertainty, dephasing, and the likelihood of photon loss. Electrical pumping can avoid this by pumping directly into the resonant state \cite{opt_v_elec_pump}. However, weakly pumped systems can complicate the process of exciting the quantum dot. For a successful recombination, the system might need to be pumped for times comparable to the emission time of the dot. This increases the probability either of multi-photon emission, on the one hand, or emitting no photons at all, on the other. For many applications, having multiple photons is more harmful than having a slightly lower single-photon probability \cite{migdall2002tailoring}. Thus, in this work, we always impose a constraint that the multi-photon emission rate is upper bounded by a small threshold.

With this in place, we improve the single-photon probability by making continuous quantum measurements on the system. The idea is to monitor the dot’s energy state continuously and determine when to stop pumping based on the information obtained. A practical approach to make this measurement is with a low-transparency quantum point contact that measures its transmission coefficient, which depends on the electric charge state of the quantum dot. This approach was studied theoretically in \cite{qpc_theory}, and experimentally verified in \cite{qpc_exp1, qpc_exp2, qpc_exp3}. We show numerically that this mechanism substantially improves the single-photon probability while also limiting the multi-photon emission rate in the weak pumping regime. This improvement exists even if we condition only on immediate measurement results, without the difficult task of processing an entire measurement record in real-time. This method is also numerically tractable since we can perform ensemble averaging with a single master equation rather than averaging over a large number of quantum trajectories. The emitter considered in this paper is a $p$-$n$ diode operated as a single-photon LED, though these techniques are applicable to other systems. We discuss the device setup and the corresponding physical processes and interactions in \secref{system_model}. We then talk about the evolution of the system in \secref{evolution}; we present a stochastic master equation and discuss the relevant parameter regime of operation. We briefly describe the evolution in the deterministic pumping case in \secref{det_pump} before introducing the threshold-based switching technique in \secref{thres_pump}. We present our numerical results in \secref{num_sims}: We first give the result in the deterministic case and then analyze the numerical performance of the threshold-based switching technique. We conclude in \secref{conclusions}.

\section{System Model} \label{system_model}

\begin{figure}
  \includegraphics[scale=.18]{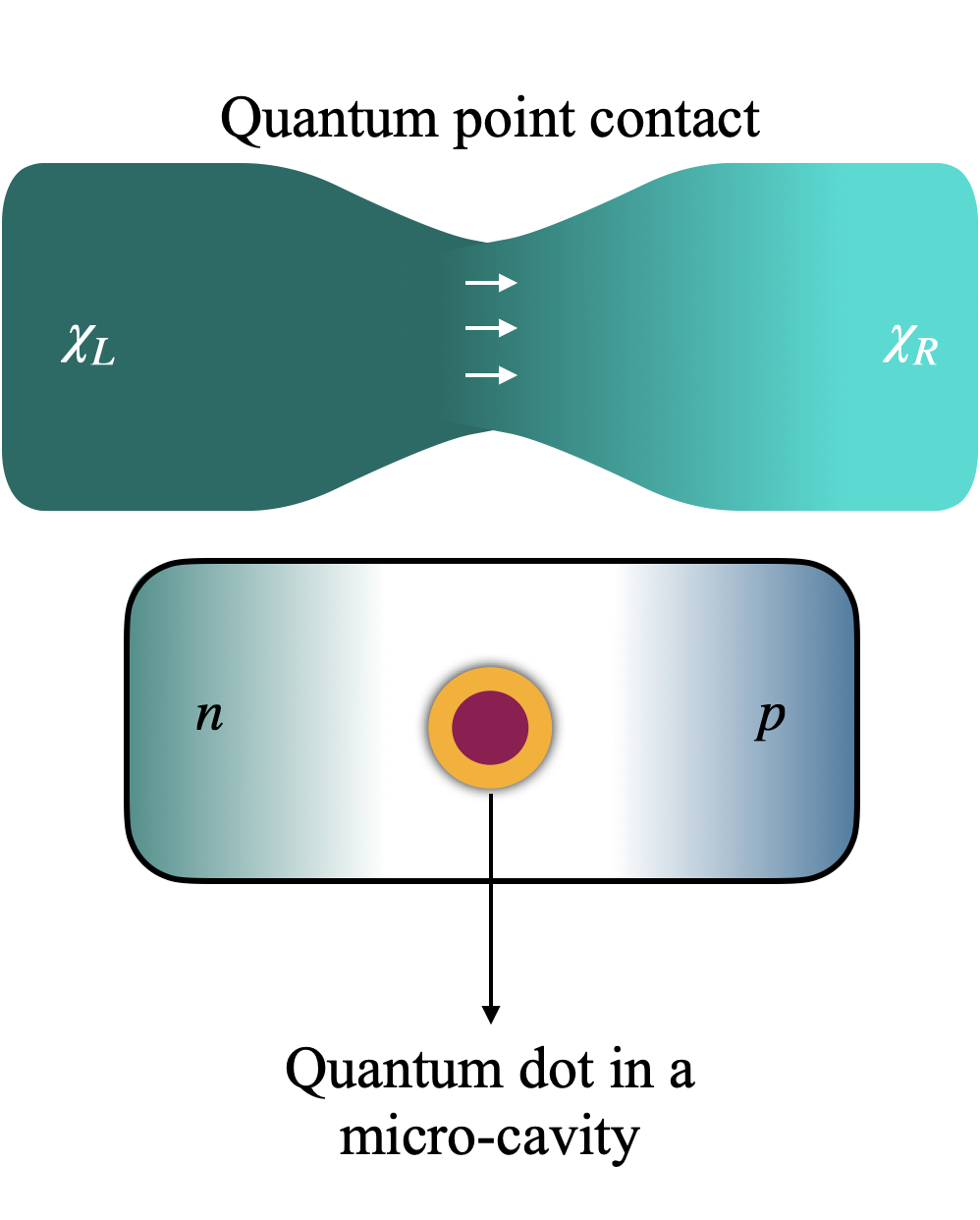}
  \caption{Quantum dot in a microcavity embedded within a $pin$ diode, whose energy state is monitor by a quantum point contact. The chemical potential in the left (right) reservoir is $\chi_L$ ($\chi_R$). The quantum dot is fabricated inside an insulator, and this is contained within an optical microcavity. The diode is biased in the forward direction, such that a single electron tunnels through from the $n$ side to the dot. Due to the Coloumb blockade effect, there is a change in the tunneling current in the point contact that can be detected.}
  \label{system}
\end{figure}

The system we are considering is a quantum dot embedded in a microcavity, which is contained within an insulator sandwiched within a \textit{pn}-junction diode. \figref{system} illustrates this. We assume that the diode is in the Coulomb blockade regime. When it is forward biased, an electron from the $n$-region tunnels through to the quantum dot, and raises its energy level. This change can be detected by using a quantum point contact as described in \cite{qpc_theory}. The energy level is reduced again when a hole tunnels from the $p$-region to the dot and recombines with the electron. This leads to the emission of a photon into the cavity, which subsequently leaks out into the waveguide.

While the photon generation process described above is intuitively straightforward, it is challenging to generate exactly one photon with high probability. Let a pumping cycle define the average tunneling duration that produced a single recombination event (and hence a single photon). The number of photons emitted then follows a Poisson counting process with the pumping cycle as the interarrival time. The conventional method of pumping lacks specific information about tunneling times. Then the best that can be done is to time the pumping a priori, either to maximize the probability of a single photon or keep the multi-photon probability below a given threshold. (These are not the same, in general.) In this work, we aim to improve this probability by using the instantaneous feedback obtained from continuous measurements.

\subsection{Formalism}

Our model comprises four subsystems: a quantum dot, an optical microcavity, an external waveguide (which we model as a bath with a single oscillator mode), and a control switch. Thus, the Hilbert space is 
\begin{equation}
    \mathcal{H} = \mathcal{H}_{dot} \otimes \mathcal{H}_{cavity} \otimes \mathcal{H}_{bath} \otimes \mathcal{H}_{control}.
\end{equation}
Our model of the dot includes two energy levels: the ground state \ket{G} and the excited state \ket{X}. The cavity and the bath contain photons and are represented by the photon number notation: $\ket{0}$, $\ket{1}$, $\ket{2}$ etc. For computational purposes, we truncate this space to 3 dimensions, with $\ket{2}$ representing the emission of 2 or more photons with probability $p(2+)$. The control switch also has two energy levels, with $\ket{1}$ ($\ket{0}$) representing the ON (OFF) state of the pumping.

The photon emission probabilities $p(0)$, $p(1)$, and $p(2+)$ can be calculated from the state of the bath subsystem. They quantitatively measure how well the system is performing. Ideally, we want the single-photon probability $p(1)$ to be close to one. Unfortunately, that is not possible for many practical parameter regimes. Moreover, certain applications are intolerant of multi-photon emission. Thus, we desire a parameter regime that maximizes $p(1)$ subject to the constraint that $p(2+) < \epsilon$, for a small $\epsilon > 0$.

\subsection{Interactions}

The total Hamiltonian can be written as $\hat{H} = \hat{H}_S + \hat{H}_I$, where we model the dot-cavity interaction by the Jaynes-Cummings Hamiltonian:
\begin{equation}
    \hat{H}_I = i\hbar g (\hat{a}^\dagger\hat{\sigma}^{-} - \hat{a}\hat{\sigma}^{+}) ,
\label{hamiltonian}
\end{equation}
where $g$ is the interaction strength, $\hat{a}^\dagger$ ($\hat{a}$) is the creation (annihilation) operator acting on the cavity mode, $\hat{\sigma}^- = \ket{G}\bra{X}$ and $\hat{\sigma}^+ = \ket{X}\bra{G}$. In the interaction picture, the entire Hamiltonian is just $\hat{H_I}$. The system is initially decoupled with the pumping ON, in the state $\ket{G,0,0,1}$.

Starting from $\ket{G,0,0,1}$, we wish to transition into the state $\ket{G,0,1,0}$, representing the dot in the ground state along with a single photon in the bath while the pumping is turned OFF. We model the system dynamics by a set of coherent and incoherent processes as follows:

\begin{itemize}
    \item \textbf{Electrical pumping}. This is the process of electron tunneling from the $n$-region to the dot under favourable bias conditions. Since this is stochastic, it is modeled as an incoherent $G \rightarrow X$ transition at a rate $\Omega$.
    \item \textbf{Dot-cavity coupling}. This is a coherent process described by the Hamiltonian \eqref{hamiltonian} with coupling strength $g$. After the electron recombines with the hole, the dot transitions back to the ground state while emitting a photon into the cavity.
    \item \textbf{Spontaneous emission}. Instead of emission into the cavity, it is possible that the excitation will be lost by emission into a different mode, or photon absorption. This incoherent process is the $X \rightarrow G$ transition at rate $\Gamma$.
    \item \textbf{Dephasing}. Once the dot is excited to $\ket{X}$, there exists Rabi oscillations between the states $\ket{X,0}$ and $\ket{G,1}$ (with the 2 modes corresponding to the dot and cavity) due to the interaction Hamiltonian. Superpositions of these states can evolve into mixtures by two processes: coupling to an external environment (whether or not the system is being monitored), or the effect of a continuous measurement (in the monitored case). These can both be modeled as dephasing at a rate $\gamma$.
    \item \textbf{Photon leakage}. This is the process of a photon transitioning from the cavity mode to the bath mode (waveguide) at a rate $\kappa$. Treating this as an incoherent process allows us to use a simple one-mode model for the bath.
    
\end{itemize}

\begin{figure*}
  \includegraphics[scale=0.5]{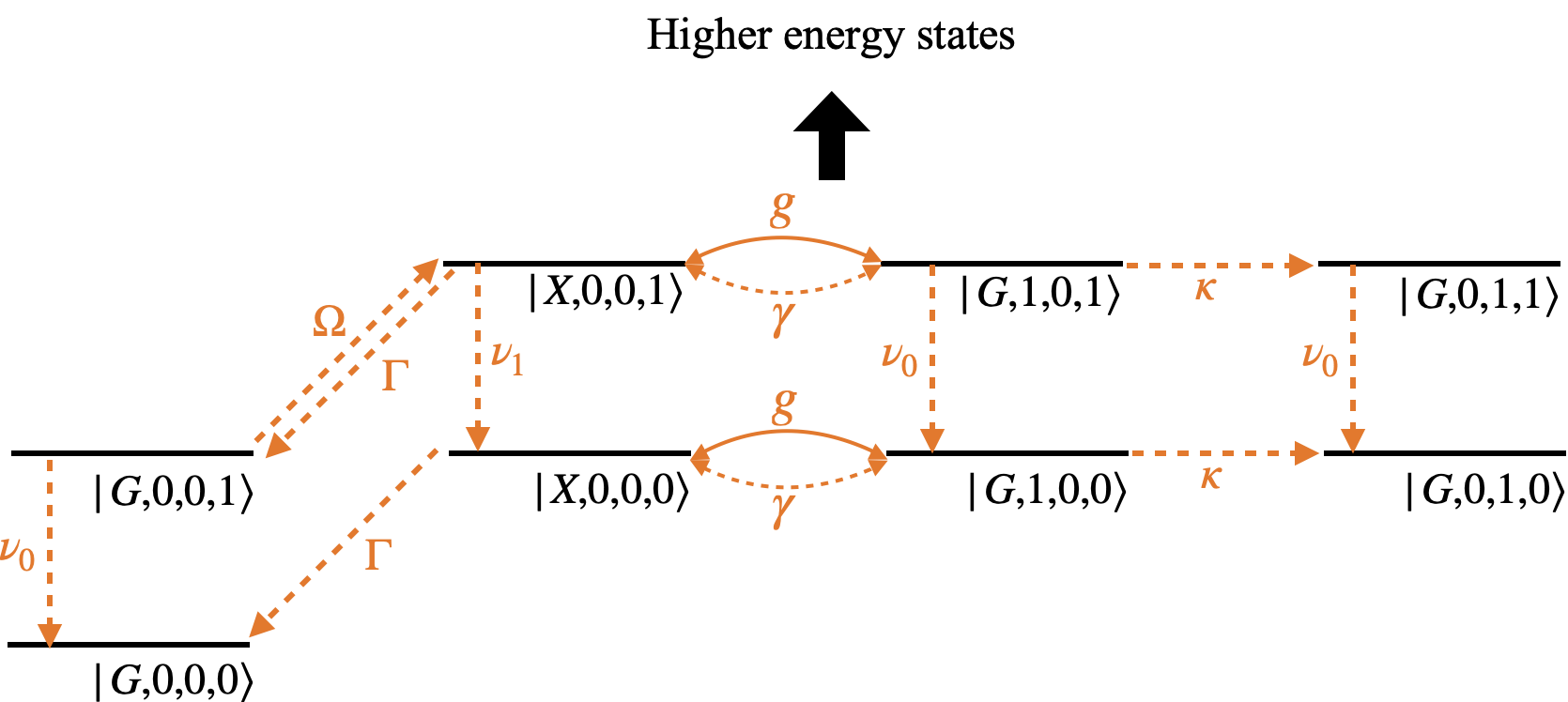}
  \caption{STATE DIAGRAM. A state of the system is described by four quantum numbers: the dot's energy level, the cavity's photon number, the external waveguide's photon number, and the ON/OFF control signal for the pumping. This diagram only includes states with at most two quanta of energy, and shows the dynamical processes that connect them. Broken arrows indicate incoherent processes, and solid arrows indicate coherent evolution. The system is initialized in the state $\ket{G,0,0,1}$, and the goal is to obtain the state $\ket{G,0,1,0}$ (corresponding to the green box) with finite pumping. Other possible final states that are undesired are $\ket{G,0,0,0}$, $\ket{X,1,0,1}$, and $\ket{X,0,1,1}$ (corresponding to the red boxes). States with higher energy have been omitted, but could be reached via the $\ket{X,1,0,1}$ and $\ket{X,0,1,1}$ states.}
  \label{statediag}
\end{figure*}

\section{System and evolution} \label{evolution}

\figref{statediag} illustrates the energy ladder and allowed transitions for the system, in which solid lines depicts coherent evolution and dashed lines represents incoherent processes. The dot will be pumped as long as the control mode is in $\ket{1}$ state, during which the system moves up the energy ladder. Depending on the dot's energy state, the control mode has different rates to make a transition from $\ket{1} \rightarrow \ket{0}$. (The rate is faster when the dot is in the state $\ket{X}$.)

Both the cavity and the bath in principle can hold infinitely many photons; but we are only interested in producing single photons with high probability. We assume that the cavity-bath coupling is stronger than the dot-cavity coupling, and hence we truncate the higher energy levels in \figref{statediag}. This means that the state $\ket{G,0,2,1}$ represents the emission of 2 or more photos, and the corresponding probability is the multi-photon probability $p(2+)$. Note that this approximation neglects some effects that could contribute in principle to the single photon probability, such as series of re-absorption and spontaneous emissions. However, the rate of leakage from the cavity to an external waveguide $\kappa$ is very large compared to the coupling $g$ between the dot and the cavity. The time the photon remains in the cavity will be of order $1/\kappa$. So the probability of reabsorption will scale like $(g/\kappa)^2$, and hence is negligibly small in the parameter regime we are considering. In our simulations, after the dot is excited, the probability of a spontaneous emission in a time interval of 0.1 is of the order $10^{-5}$. Hence, the neglected effects should have little impact on the single photon probability estimates.

In general, quantum dynamics are described by quantum dynamical maps,
\begin{equation}
    \rho(t) = \Phi_t[\rho(0)] ,
    \label{dyn_map}
\end{equation}
where $\Phi_t$ is a completely positive trace-preserving (CPTP), time-dependent map with $\Phi_0 = I$. In the case of a Markovian quantum master equation, the dynamical map superoperator $\Phi_t$ can be written as
\begin{equation}\label{eq:lind_evol}
    \Phi_{t}=T \exp \left[\int_{0}^{t} \mathcal{L}_{\tau} d \tau\right],
\end{equation}
where $T$ denotes Dyson time ordering and $\mathcal{L}_{\tau}$ is the time-dependent Gorini–Kossakowski–Sudarshan–Lindblad (GKSL) generator \cite{qdyn},
\begin{equation}
    \mathcal{L}_{\tau} \rho=-\frac{i}{\hbar}[H, \rho]+\sum_{k} \alpha_{k}\left(L_{k} \rho L_{k}^{\dagger}-\frac{1}{2}\left\{L_{k}^{\dagger} L_{k}, \rho\right\}\right),
\end{equation}
where $\{L_k\}$ are the Lindblad operators with corresponding rates $\alpha_k \geq 0$. Therefore, Eq.~\eqref{dyn_map} becomes
\begin{equation}
    \frac{d}{dt} \rho(t) = \mathcal{L} \rho(t).
\end{equation}

In our case, the quantum map does not change with time. Hence, we have a time-independent GKSL generator and Eq. \eqref{eq:lind_evol} becomes $\Phi_{t}=e^{\mathcal{L}t}$. We can then evolve the state $\rho_0$ by solving the vector differential equation,
\begin{equation}
    \rho(t) = e^{\mathcal{L}t}\rho(0) = \sum_j c_j e^{\lambda_j t} \textbf{v}_j,
    \label{vec_diff_eq}
\end{equation}
where $\textbf{v}_j$ is an eigenvector of $\mathcal{L}$ with the corresponding eigenvalue $\lambda_j$ and coefficient $c_j$. As $t \rightarrow \infty$, all the components in Eq.~\eqref{vec_diff_eq} with $\mathrm{Re}\{\lambda_j\} < 0$ decay to $0$, and only the stable components (with $\mathrm{Re}\{\lambda_j\} = 0$) remain. These states represent the asymptotic states that result from the quantum evolution.

In our case, we wish to pump the quantum dot (which can be thought of as an evolution in accordance with GKSL generator $\mathcal{A}$) for a certain amount of time (say, $T_s$), then turn off the pumping and continue the evolution (now with different evolution dynamics, in accordance with GKSL generator $\mathcal{B}$). The quantum state at time $T_s$ can be expressed as
\begin{equation}
    \rho(T_s) = \sum_k \beta_k \textbf{u}_k = \sum_j \alpha_j e^{\lambda_j T_s} \textbf{v}_j,
    \label{evol_transit}
\end{equation}
where \textbf{v} (respectively, \textbf{u}) are the eigenvectors of $\mathcal{A}_{\tau}$ (respectively, $\mathcal{B}_\tau$), with eigenvalues $\lambda$ ($\lambda'$) and vector coefficients $\alpha$ ($\beta$). At times before $T_s$ the system evolves using the first set of eigenvectors and eigenvalues; after $T_s$, the evolution continues using the new eigenvectors and eigenvalues. (One must therefore change basis at time $T_s$.)

In the next section, we explain the evolution under the deterministic pumping scenario; in the section after, we focus on the main proposal of this paper: the threshold-based switching technique with the aid of continuous measurements.

\subsection{Deterministic pumping}\label{det_pump}

We introduce the system evolution in the simplest case by giving a brief description in the deterministic pumping scenario. This is the case with no information about the pumping duration. Since there are no measurements, the system evolves deterministically, which corresponds to zero dephasing, $\gamma = 0$ (though in general, of course, there could be dephasing due to environmental interactions; for the moment we assume this is negligible on these time scales). Also since there is no way to know the dot's energy, the control subsystem is trivial and can be dropped from the system. Then the evolution of the system can be described by the following Lindblad master equation:
\begin{equation}
    d\rho = -\frac{i}{\hbar} [\hat{H}, \rho] dt + \left(\Omega \mathcal{D}[\hat{\sigma}^+] + \Gamma\mathcal{D}[\hat{\sigma}^{-}] + \kappa \mathcal{D}[\hat{a}\hat{b}^\dag]\right)\rho dt,
\label{det_sme}
\end{equation}
where $a^\dagger$ (respectively, $b^\dagger$) represents the creation operator of the cavity (bath), and $\mathcal{D}$ is a superoperator (a Lindblad term) defined as:
\begin{equation}
\mathcal{D}[\hat{A}]\rho = \hat{A}\rho\hat{A}^\dagger
- \frac{1}{2}\{\hat{A}^\dagger\hat{A}, \rho\}.
\end{equation}
The first Lindblad term represents pumping of the quantum dot, with rate $\Omega$; the second represents spontaneous emission, with rate $\Gamma$; and the third represents leakage from the cavity to the external waveguide (bath) with rate $\kappa$. The coupling between the quantum dot and the cavity is contained in the Hamiltonian of form \eqref{hamiltonian}, and has coupling strength $g$. In deterministic evolution, the system evolves according to this master equation up to a fixed time $T_s$, at which point the pumping is turned off, and the evolution continues with $\Omega=0$. The changeover from one evolution equation to another is done using the basis change in Eq.~\eqref{evol_transit} at time $T_s$.

\subsection{Threshold-based switching}\label{thres_pump}

As mentioned earlier, deterministic pumping may not lead to the efficient generation of single photons due to the stochastic nature of the pumping cycle. We thus consider a technique that weakly measures the energy of the quantum dot at each instant and uses that as feedback to control the pumping. To obtain the energy of the dot, we measure the projection operator onto the excited state of the dot, denoted by $\mathcal{\hat{P}}_X$. This measurement can be made using a quantum point contact, as shown in \figref{system}. Reference \cite{qpc_theory} considered a similar model and described the evolution in terms of a stochastic master equation in the weak measurement regime (diffusive limit). The measurement operator they considered was proportional to the occupation of the quantum dot, truncated to a 2-dimensional subspace, which in turn is proportional to $\mathcal{\hat{P}}_X$. References \cite{qpc_exp1, qpc_exp2, qpc_exp3} experimentally showed the possibilities of such a measurement. The measurement outcome then should be a noisy estimate of the population in the excited state. This measurement outcome is then used to control the pumping by simply ``flagging'' if the energy exceeds a given threshold $\tau$ or not. To mitigate the consequences of a false-negative flag, the pumping is turned off after a maximum pumping time $T_s$ even if the threshold was never exceeded. Just as in the deterministic case, the changeover from one evolution equation to another is done using the basis change in Eq.~\eqref{evol_transit} at time $T_s$. The threshold $\tau$ and maximum time $T_s$ are hyperparameters that are chosen to optimize the performance of this protocol. This threshold-based approach is not optimal; by only switching using the measurement output at one time, we neglect the possibility of extracting information from the entire measurement record. However, there are reasons why we might prefer such a threshold-based protocol. It is much simpler, with lower latency, than storing and processing an entire measurement record, which may not be practical in real-time. Moreover, solving a single master equation is sufficient to estimate the performance of this method for a given set of parameters, rather than averaging over many stochastic quantum trajectories.

The output signal obtained from the continuous weak measurement (in rescaled units) is given by
\begin{equation}
    I(t) = \langle\hat{\mathcal{P}}_X\rangle(t) + \beta \frac{dW_t}{dt},
\label{measurementOutput}
\end{equation}
where $\beta = (\eta\gamma)^{-1/2}$. Here, $\gamma$ is the measurement rate, $\eta$ is the measurement efficiency and $dW_t$ is the Wiener process with variance $dt$. So $\mathbb{E}[dW_t] = 0$ and $\mathbb{E}[dW_tdW_s] = \delta(t-s)dsdt$, and $\langle \cdot \rangle=\text{Tr}[\cdot\rho(t)]$ is the quantum expectation.

As described in \secref{system_model}, the pumping of the dot is regulated by the control mode. In the approach we are considering, the control mode evolves based only on the current, weakly measured state of the quantum dot; it does not have access to the measurement record at earlier times. The goal is to stop pumping as soon as the dot is excited to a higher energy level, which involves switching the control mode from $\ket{1} \rightarrow \ket{0}$. This switching is based on the measurement output in Eq.~\eqref{measurementOutput} exceeding a threshold $\tau$ or not. This can be modeled in the master equation as an incoherent process in which the switching of the control mode occurs at a rate that is dependent on the dot's energy state: if the dot is in the ground state, the rate is lower than when the dot is in a higher energy level.

To calculate these rates and their dependence on $\tau$, first consider the output signal averaged over a succession of intervals of size $\Delta t$. Since it is a Gaussian random variable, the probability of the dot's energy exceeding a threshold $\tau$ is given by
\begin{equation}
    \begin{split}
        p(\bar{I} > \tau) &= \int_{\tau}^{\infty} p(\bar{I}) \,d\bar{I} \\
        &\approx \frac{\exp{\left(-\eta\gamma\Delta t \left(\frac{\tau}{\Delta t}-\mu\right)^2/2\right)}}{\sqrt{2\pi\eta\gamma \Delta t}(\frac{\tau}{\Delta t}-\mu)}.
    \end{split}
    \label{prob}
\end{equation}
where $\mu=\langle \hat{\mathcal{P}}_X \rangle$. A detailed derivation of the above approximation is given in Appendix~A.
Then the rate at which we stop pumping is equal to
\begin{equation}
    \nu_{\mu} = \frac{p(\bar{I} > \tau)}{\Delta t}.
\end{equation}
Clearly, this quantity depends on the quantum dot's energy state. Moreover, the rates to stop pumping corresponding to the quantum dot being completely in the ground state and the excited state are related to each other (as derived in Appendix~B):
\begin{equation}
    \nu_0 \approx \nu_1 \exp\left[-\sqrt{(2\eta\gamma\Delta t)\ln\left(\frac{1}{\nu_1\Delta t \sqrt{2\pi\eta\gamma\Delta t}}\right)}\right].
\end{equation}

Finally, the evolution of the system under the threshold-based switching approach is described by the Lindblad master equation \eqref{sme}:
\begin{widetext}
\begin{equation}
    d\rho = -\frac{i}{\hbar} [\hat{H}, \rho] dt + \left(
    \Omega \mathcal{D}[\hat{\sigma}^+\hat{\xi}] + 
    \Gamma\mathcal{D}[\hat{\sigma}^{-}] + 
    \kappa \mathcal{D}[\hat{a}\hat{b}^\dag] + 
    \gamma \mathcal{D}[\hat{\mathcal{P}}_X] + 
    \nu_1 \mathcal{D}[\hat{c}\;\hat{\mathcal{P}}_X] +
    \nu_0 \mathcal{D}[\hat{c}\;(I-\hat{\mathcal{P}}_X)]
 \right)\rho dt ,
\label{sme}
\end{equation}
\end{widetext}
where $\hat{\xi}=|1\rangle_{\textrm{control}}\langle 1| $ is non-zero only when the control is not $|0\rangle$, and $\hat{c}$ is the annihilation operator for the control signal.
\vspace*{-5px}
\subsection{Parameter regime}

The spontaneous emission rate $\Gamma$ should be much smaller than the pumping rate $\Omega$ that translates to a favourable forward bias of the diode. Electrical pumping typically corresponds to weak coupling between the dot and the cavity (small $g$). We also assume that the photon leakage rate $\kappa$ is dominant, so that photons in the cavity are promptly transferred to the external waveguide. (In practice, there are some trade-offs between these parameters---high $\kappa$ limits the quality factor $Q$ of the cavity, which reduces $g$ and may prevent us from making $\Gamma$ as small as we might like. We do not explicitly model these trade-offs in this work.)

Since this technique involves continuous measurements of the dot that are generally slower than the coherent evolution, we assume that the dot-cavity coupling strength $g$ is comparable to the pumping rate, to ensure that there is sufficient time to make a decision to turn off the pumping. However, in the case where $g \ll \Omega$, this approach still works by turning on the pumping for a time much less than the emission time scale. In this regime, feedback gives little advantage.

Considering the above constraints, the following parameter regime is where we expect this technique to give a significant advantage in performance:
\begin{equation}
    \Gamma \ll \nu_0 < \Omega, g < \nu_1 \le \gamma, \kappa.
\end{equation}

\section{Numerical simulations} \label{num_sims}
Above, we presented an approach to improve the probability of emitting single photons using threshold-based switching. It is a simple method, since it only requires comparing the instantaneous measurement output to a fixed threhshold, and numerically tractable, since we can perform ensemble averaging with a single master equation. The decision to turn off the pumping is implicit within the master equation. Moreover, the controllable parameters $\nu_1$ (or, equivalently, $\tau$), $\gamma$ and stopping time $T_s$ can be set such that we never perform worse than the deterministic (open-loop) case for the same fixed parameter values $\Omega$ $g$ and $\kappa$ (assuming $\Gamma$ is very small). As we will see, in some parameter ranges we do significantly better.

We first benchmark the results obtained by the deterministic approach, and then compare those with the ones obtained by the threshold-based switching technique. We optimize the parameters to find the maximum attainable single-photon probability while also limiting the multi-photon probability. We expect that this method will be particularly useful with relatively high measurement strength. Although that translates to a high dephasing rate, we show that it is possible to choose a parameter regime that still out-performs the deterministic case.

\subsection{Time scale}\label{time_scale}

Dynamics of our system depend on a multitude of Lindblad rates--- $\Omega, g, \Gamma, \gamma$, and $\kappa$. We rescale all the parameters with respect to the rate of emission into the waveguide $\kappa$. This establishes the dimensionless units for the simulation. In these units, the parameter value ranges we are particularly interested in are:
\begin{flalign}\label{parvals}
        \Omega, g &\in [0.01, 0.1],\\
        \gamma &\in \{0.1, 1.0, 10.0\},\\
        \kappa &= 1.0, \\
        \Gamma &\ll 0.1.
\end{flalign}

\subsection{Parameter optimization}\label{par_opt}

\begin{figure}[!h]
    \begin{tikzpicture}
        \begin{axis}[
            xlabel = \(T_s\),
            ylabel = {\(\text{probability}\)},
            xmin=0, xmax=100,
            ymin=0.0, ymax=1.0,
            xtick={0, 20, 40, 60, 80, 100},
            ytick={0,0.2,0.4,0.6,0.8,1.0},
            legend pos=south east,
            xmajorgrids=true,
            ymajorgrids=true,
            grid style=solid,
            line width=0.7 pt
        ]
        \addplot[
            solid,
            color=blue,
            decoration={text along path,
            text={{p(0)}{}},
            text align={align=center, left indent=2cm},
            raise=4pt
            },
            postaction={decorate}             
            ]
            table[col sep=comma] {det_evol_p0.dat};
        \addplot[
            solid,
            color=orange,
            decoration={text along path,
            text={{p(1)}{}},
            text align={align=right, right indent=0.8cm},
            raise=4pt
            },
            postaction={decorate}             
            ]
            table[col sep=comma] {det_evol_p1.dat};
        \addplot[
            solid,
            color=olive,
            decoration={text along path,
            text={{p(2+)}{}},
            text align={align=right, right indent=0.8cm},
            raise=4pt
            },
            postaction={decorate}             
            ]
            table[col sep=comma] {det_evol_p2.dat};
        \end{axis}
    \end{tikzpicture}
  \caption{DETERMINISTIC EVOLUTION. We plot the asymptotic photon probabilities——$p_0$, $p_1$ and $p_{2+}$ as a function of the stopping time $T_s$. The parameter values for this example are $\Omega=0.1, g=0.1, \Gamma=0.001$ and $\kappa=1.0$. The system begins in the state \ket{G,0,0}. When $T_s=0$, we do not pump at all, hence $p_0(0)=1$ asymptotically. As we increase $T_s$, the system undergoes non-trivial evolution according to Eq.~\eqref{det_sme}; $p_1$ reaches its maximum value and then starts to decrease, while $p_{2+}$ increases monotonically with $T_s$ as the system keeps emitting more photons to the waveguide.}
  \label{detSol}
\end{figure}
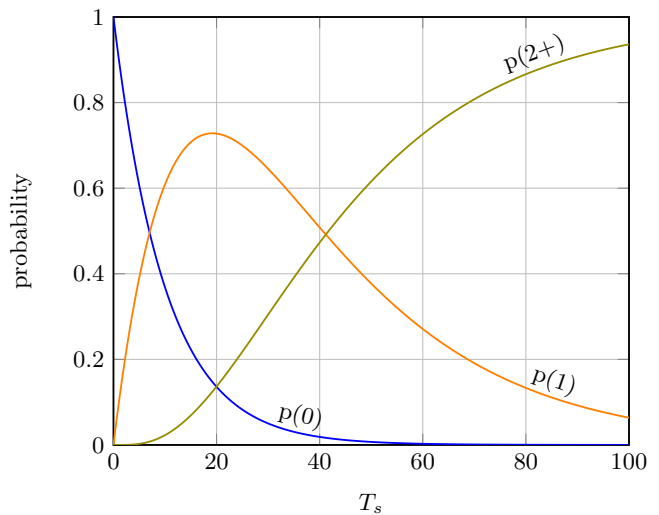

Here, we consider evolution with deterministic pumping for a fixed time $T_s$. Our goal is to maximize the single-photon probability while keeping the multi-photon probability below a certain limit $\epsilon$. (This optimization approach can be generalized to the threshold-based switching technique with some additional constraints, as we will see.) To evolve the system deterministically, we numerically integrate $d\rho$ as defined in Eq.~\eqref{det_sme} using the $4^\textrm{th}$-order Runge-Kutta method. \figref{detSol} shows the asymptotic probabilities (i.e., at long times after all emissions have occurred) of emitting photons as a function of the stopping time $p_i(Ts)$ where $i \in \{0,1,2+\}$ corresponding to zero-photon, single-photon and multi-photon probabilities. The figure shows a particular set of Lindblad rates, but the qualitative behavior is similar for all parameter values. Let us define two times: $t'$, which is the time when $p_2(t')=\epsilon$, and $t''=\argmax_t p_1(t)$. We observe that, depending on the parameter regime, we are in one of 2 cases: $t' < t''$ or $t' \geq t''$. In the former case, we have not reached the maximum attainable $p_1$, but we have already reached the maximum tolerable $p_2$. In the latter case, we have already crossed the maximum attainable $p_1$. Hence, the optimal time to stop pumping is the minimum of $t'$ and $t''$.

Suppose we wish to keep $p_2(T_s)\leq\epsilon$. Then for each set of parameters $\Omega, g$, we calculate the following:
\begin{flalign}
    t_\epsilon &\equiv t' \;\; \textrm{where} \;\; p_2(t')=\epsilon , \\
    t_m &\equiv \argmax_t p_1(t) , \\
    t_\textrm{opt} &= \min(t_\epsilon, t_m).
\end{flalign}
Then the maximum attainable single-photon probability while keeping the multi-photon probability below $\epsilon$ is $p_1(t_\textrm{opt})$. In the deterministic case, there is an unavoidable trade-off between $p_2$ and $p_1$; if we want to limit the former, we cannot maximize the latter. Continuous measurements and feedback can ease this trade-off.

Optimizing in the threshold-based switching technique imposes additional constraints on $\nu_1$ and $\gamma$. In this paper, we only considered a discrete set of values of $\gamma$, thus easing the optimization procedure, but numerically optimizing over a full range of values could be done. We analytically find the asymptotic probabilities $p_i(\nu_1, T_s)$, for $i\in\{0,1,2+\}$, as a function of $\nu_1$ and $T_s$. Then following a similar approach as above, for each set of parameters $\Omega, g, \gamma$, we calculate the following:
\begin{flalign}
    t_\epsilon(\nu_1) &\equiv t' \;\; \textrm{where} \;\; p_2(\nu_1, t')=\epsilon , \\
    p^{(1)}_{\epsilon} &\equiv \max_{\nu_1} p_1(\nu_1, t_\epsilon(\nu_1)) , \\
    \nu_1', t' &= \argmax_{\nu_1, t} p_1(\nu_1, t) , \\
    p_1' &= p_1(\nu_1', t') , \\
    p_2' &= p_2(\nu_1', t').
\end{flalign}
Then the maximum attainable single-photon probability while keeping the multi-photon probability below $\epsilon$ is 
\begin{equation}
p_1^{*}=
\begin{cases}
    p_1', & \text{if } p_2' \leq \epsilon \\
    p^{(1)}_{\epsilon}, & \text{otherwise}
\end{cases}
\end{equation}

\subsection{Effect of pumping rate}

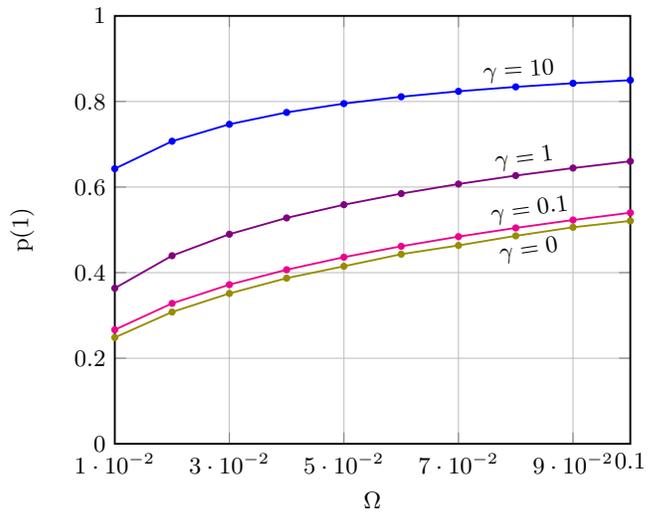
\begin{figure}[!ht]

    \begin{tikzpicture}
        \begin{axis}[
            xlabel = \(\Omega\),
            ylabel = {\(\text{p}(1)\)},
            xmin=0.01, xmax=0.1,
            ymin=0.00, ymax=1.0,
            legend pos=south east,
            xmajorgrids=true,
            ymajorgrids=true,
            grid style=solid,
            line width=1 pt,
            scaled x ticks=false, 
            scaled y ticks=false, 
            x tick label style={
              /pgf/number format/.cd,
              fixed,              
              precision=2         
            },
            y tick label style={
              /pgf/number format/.cd,
              fixed,              
              precision=2         
            }
          ]
        ]
        \addplot[
            solid,
            color=olive,
            mark=*,
            mark size=1.0pt,
            decoration={text along path,
            text={{$\gamma=0$}{}},
            text align={align=right, right indent=1cm},
            raise=-8pt
            },
            mark options={decoration={name=none}},
            postaction={decorate}            
            ]
            coordinates {
            (0.01, 0.248452) (0.02, 0.307969) (0.03, 0.351341) (0.04, 0.387079) (0.05, 0.414827) (0.06, 0.443039) (0.07, 0.463575) (0.08, 0.485954) (0.09, 0.505979) (0.1, 0.520906)
            };
        \addplot[
            solid,
            color=magenta,
            mark=*,
            mark size=1.0pt,
            decoration={text along path,
            text={{$\gamma=0.1$}{}},
            text align={align=right, right indent=0.8cm},
            raise=4pt
            },
            mark options={decoration={name=none}},
            postaction={decorate}             
            ]
            coordinates {
            (0.01, 0.266553) (0.02, 0.328123) (0.03, 0.371905) (0.04, 0.406787) (0.05, 0.436119) (0.06, 0.461582) (0.07, 0.484157) (0.08, 0.504475) (0.09, 0.522968) (0.1, 0.539949)
            };
        \addplot[
            solid,
            color=violet,
            mark=*,
            mark size=1.0pt,
            decoration={text along path,
            text={{$\gamma=1$}{}},
            text align={align=right, right indent=1cm},
            raise=4pt
            },
            mark options={decoration={name=none}},
            postaction={decorate}            
            ]
            coordinates {
            (0.01, 0.363623) (0.02, 0.439579) (0.03, 0.489791) (0.04, 0.527873) (0.05, 0.558706) (0.06, 0.584659) (0.07, 0.607073) (0.08, 0.626794) (0.09, 0.644387) (0.1, 0.660253)
            };
        \addplot[
            solid,
            color=blue,
            mark=*,
            mark size=1.0pt,
            decoration={text along path,
            text={{$\gamma=10$}{}},
            text align={align=right, right indent=1cm},
            raise=4pt
            },
            mark options={decoration={name=none}},
            postaction={decorate}             
            ]
            coordinates {
            (0.01, 0.64288) (0.02, 0.707055) (0.03, 0.746703) (0.04, 0.774401) (0.05, 0.795024) (0.06, 0.811003) (0.07, 0.823725) (0.08, 0.834059) (0.09, 0.84258) (0.1, 0.849691)
            };
        \end{axis}
    \end{tikzpicture}
  \caption{Maximum attainable single photon probability as a function of pumping rate $\Omega$ for $g=0.1$. The multi-photon emission probability $p(2) \leq 1\%$. There are 4 cases corresponding to different measurement strengths $\gamma$: $(i)$ deterministic pumping (no measurement), $(ii)$ low measurement rate ($\gamma=0.1$), $(iii)$ intermediate measurement rate ($\gamma=1$), $(iv)$ high measurement rate ($\gamma=10$). High measurement rate $(iv)$ leads to the best performance, while the low measurement rate $(ii)$ performs the worst among the measurement-based schemes, although still marginally better than the deterministic case.}
  \label{pump_plt}
\end{figure}

It is straightforward to see that the single photon probability increases as we increase the pumping rate. But for a weakly pumped system, such as the electrical pumping considered in this work, we wish to have a good $p_1$ even in the low pumping rate regime. The threshold-based switching technique addresses this issue by ``flagging'' the control signal as soon as the dot is excited to a higher energy level. This in turn allows the system to promptly switch off the pumping. In \figref{pump_plt}, we plot the maximum attainable single photon probability as a function of pumping rate $\Omega$. We set the maximum tolerable level of multi-photon emission $p_2$ here and in further simulations to $1\%$. There are 4 cases corresponding to different measurement strengths $\gamma$: $(i)$ deterministic pumping (no measurement), $(ii)$ low measurement rate ($\gamma=0.1$), $(iii)$ intermediate measurement rate ($\gamma=1$), and $(iv)$ high measurement rate ($\gamma=10$); high measurement rate $(iv)$ leads to the best performance, while low measurement rate $(ii)$ performs the worst among the measurement-based schemes, although still marginally better than the deterministic case.  A lower pumping rate means that the electron tunneling events are spread out in time, thus giving the threshold-based switching technique a long time to make the right decision. For higher rates, this time window is diminished, leading to a lesser performance improvement.

\vspace*{-10px}
\subsection{Effect of coupling strength}

\begin{figure}[!ht]
    \centering
    \begin{tikzpicture}
        \begin{axis}[
            xlabel = \(g\),
            ylabel = {\(\text{p}(1)\)},
            xmin=0.02, xmax=0.1,
            ymin=0.00, ymax=1.0,
            legend pos=south east,
            xmajorgrids=true,
            ymajorgrids=true,
            grid style=solid,
            line width=0.7 pt,
            scaled x ticks=false, 
            scaled y ticks=false, 
            x tick label style={
              /pgf/number format/.cd,
              fixed,              
              precision=2         
            },
            y tick label style={
              /pgf/number format/.cd,
              fixed,              
              precision=2         
            }
          ]
        ]
        \addplot[
            solid,
            color=red,
            mark=asterisk,
            decoration={text along path,
            text={Deterministic},
            text align={align=right, right indent=0.3cm},
            raise=4pt
            },
            mark options={decoration={name=none}},
            postaction={decorate}
            ]
            coordinates {
            (0.02, 0.848042) (0.03, 0.792325) (0.04, 0.731856) (0.05, 0.681757) (0.06, 0.638864) (0.07, 0.603691) (0.08, 0.572662) (0.09, 0.5434) (0.1, 0.520906)
            };
        \addplot[
            solid,
            color=blue,
            mark=diamond*,
            decoration={text along path,
            text={Threshold-based},
            text align={align=right, right indent=0.3cm},
            raise=4pt
            },
            mark options={decoration={name=none}},
            postaction={decorate}
            ]
            coordinates {
            (0.02, 0.850489) (0.03, 0.84345) (0.04, 0.841092) (0.05, 0.840067) (0.06, 0.839531) (0.07, 0.840823) (0.08, 0.844442) (0.09, 0.847095) (0.1, 0.848729)
            };
        \end{axis}
    \end{tikzpicture}
    \caption{Maximum attainable single photon probability as a function of coupling rate $g$, comparing $(i)$ deterministic pumping to the $(ii)$ threshold-based switching technique.  The multi-photon emission probability $p(2) \leq 1\%$. Plots are optimized numerically over $\Omega$, $\nu_1$, $\gamma$, and $T_s$ in accordance with the parameter ranges in \eqref{parvals}.}
    \label{fig:coup_plt}

\end{figure}
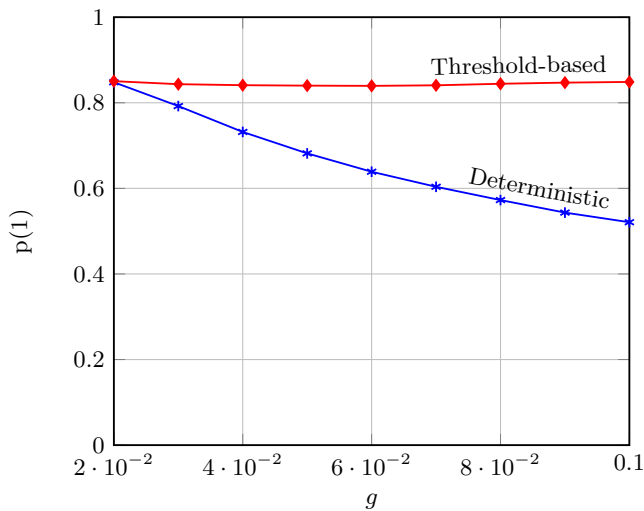
At low coupling strengths between the dot and the cavity, the threshold-based switching method is not particularly useful. It does not perform appreciably better than the deterministic case. However, as the coupling strength is increased, a significant fraction of the population in the $\ket{X,0,0}$ state is transferred to the $\ket{G,1,0}$ state very quickly. If the pumping is still turned on, more photons are added to the cavity, reducing the single photon probability. However, if the pumping is turned off just as the dot is excited to the excited state, the coupling rate will not strongly affect the single photon probability. This is illustrated in \figref{fig:coup_plt}. In the low-coupling regime, both methods produce single photons with about $85\%$ probability. As we increase the coupling strength, the deterministic method drops the single photon rate rapidly, while the threshold-based approach keeps it at $85\%$. Again, the multi-photon emission probability is capped at $1\%$.

\section{Conclusions}\label{conclusions}

Threshold-based switching is a simple yet powerful technique that can significantly improve the single-photon probability without increasing the multi-photon probability, by continuously measuring a quantum dot's energy and using feedback to control when to stop pumping.  We modeled the system by a stochastic master equation that includes post-measurement operations. This technique is useful particularly at low pumping rates and high coupling rates, where it outperforms the deterministic case. However, in regions of strong pumping or low coupling, the improvement over the deterministic pumping (no monitoring at all) is only marginal. Numerical simulations showed that even a simple threshold-based feedback scheme using continuous measurements can improve performance over deterministic (open-loop) pumping in certain parameter regimes, and always performs at least as well.

\begin{acknowledgments}
The authors would like to thank Yi-Hsieng Chen, Namit Anand, Christopher Sutherland and Prithviraj Prabhu for useful discussions. This work is supported by the National Science Foundation under Award No.~1911089.
\end{acknowledgments}

\bibliography{references}

\appendix

\section{Cumulative distribution of measurement current} \label{sec:cdf_meas}

The measurement current (in re-scaled units) is given by
\begin{equation}
    i(t) = \langle \hat{\mathcal{P}}_X \rangle (t) + \sqrt{\frac{1}{\eta\gamma}}\frac{dW_t}{dt}.
\end{equation}
Let $\mu \equiv \hat{\mathcal{P}}_X$ be the projection onto the excited state of the quantum dot, and is the observable we are measuring; $\gamma$ is the measurement strength with efficiency $\eta$. Then,
\begin{equation}
         \bar{i}(t) = \int_t^{t+\Delta t} i(t)dt. \\
\end{equation}
But $dW_t \sim \mathcal{N}(0, \Delta t)$. Hence,
\begin{equation}
\begin{aligned}
    \bar{i}(t) &\sim \mu \Delta t+\frac{1}{\sqrt{\eta \gamma}} \mathcal{N}(0, \Delta t), \\
               &\sim \mathcal{N}\left(\mu \Delta t, \frac{\Delta t}{\eta \gamma}\right). \\
    \end{aligned}
\end{equation}
Now the probability that the measurement current exceeds a threshold $\tau$ (in other words, the probability that the population of the quantum dot in the excited state exceeds $\tau$) is
\begin{equation}
\begin{aligned}
    p(\bar{i} > \tau) &= \int_{\tau}^{\infty} p(\bar{i}) d\bar{i}, \\
                      &= \frac{1}{\sqrt{2\pi\Delta t/\eta\gamma}}\int_\tau^{\infty} \exp \left(\frac{-(\bar{i}-\mu\Delta t)^2}{2\Delta t/\eta\gamma}\right), \\
                      &= \frac{1}{2} \text{erfc}\left\{\sqrt{\eta\gamma\Delta t/2} \left(\frac{\tau}{\Delta t}-\mu\right) \right\}.
\end{aligned}
\end{equation}
Using the series expansion of the complementary error function, the above equation can be written in the first order approximation as
\begin{equation}
    p(\bar{i} > \tau) \approx \frac{\exp{\left(-\eta\gamma\Delta t \left(\frac{\tau}{\Delta t}-\mu\right)^2/2\right)}}{\sqrt{2\pi\eta\gamma \Delta t}(\frac{\tau}{\Delta t}-\mu)}
\end{equation}
which is given in \eqref{prob}. In this paper we set $\eta=1$.

\section{Lindblad rates to turn control signal off}\label{app:rates}

The rate to stop pumping depends on the quantum dot's energy state $\mu$, and is given by
\begin{equation}
\begin{aligned}
    \nu_\mu &\equiv \frac{p(\bar{i}>\tau)(\mu)}{\Delta t}, \\
            &=      \frac{1}{\sqrt{2\pi\eta\gamma \Delta t}(T-\mu)\Delta t} \exp{\left(\frac{-\eta\gamma\Delta t (T-\mu)^2}{2}\right)},
\end{aligned}
\end{equation}
where $T \equiv \tau/\Delta t \gg 1$. Clearly, the above equation depends on the quantum dot's energy state $\mu$. In our case, we limit the quantum dot to be either in the ground state or the first excited state, corresponding to $\mu=0$ or $\mu=1$ respectively. We then compute the corresponding rates:
\begin{equation}
\begin{aligned}
    \nu_0 &= \frac{1}{\sqrt{2\pi\eta\gamma \Delta t}T\Delta t} \exp{\left(\frac{-\eta\gamma\Delta t T^2}{2}\right)}, \\
    \nu_1 &= \frac{1}{\sqrt{2\pi\eta\gamma \Delta t}(T-1)\Delta t} \exp{\left(\frac{-\eta\gamma\Delta t (T-1)^2}{2}\right)}.
\end{aligned}
\label{appeq:rates}
\end{equation}
Taking the ratio of both the rates, we obtain
\begin{equation}
    \frac{\nu_1}{\nu_0} = \frac{T}{T-1} \exp \left(-\eta\gamma\Delta t (T-0.5)\right).
\end{equation}
Since $T\gg1$, the equation can be approximated as
\begin{equation}
\label{appeq:approx_rates}
    \nu_0 \approx \nu_1 e^{-\eta\gamma\Delta t T}.
\end{equation}
Combining equations (\ref{appeq:rates}) and (\ref{appeq:approx_rates}), we obtain
\begin{equation}
    \frac{1}{T\Delta t \sqrt{2\pi\eta\gamma\Delta t}} \exp \left(\frac{-\eta\gamma\Delta t T^2}{2}\right) \approx \nu_1 e^{-\eta\gamma\Delta t T}.
\end{equation}
Taking natural log on both the sides and rearranging the terms,
\begin{equation}
    \frac{-\eta\gamma\Delta t T^2}{2} - \ln{T} -\ln\left[\nu_1\Delta t \sqrt{2\pi\eta\gamma\Delta t}\right] \approx 0.
\end{equation}
But $\ln{T}$ is small when compared to the term in $\mathcal{O}(T^2)$ and can be ignored. Hence we have
\begin{equation}
    T^2 - 2T + \frac{2\ln\left[\nu_1\Delta t \sqrt{2\pi\eta\gamma\Delta t}\right]}{\eta\gamma\Delta t} \approx 0.
\end{equation}
Solving the above quadratic equation,
\begin{equation}
    \begin{aligned}
        T &= 1 \pm \sqrt{1 + \left(\frac{2}{\eta\gamma\Delta t}\right)\ln\left(\frac{1}{\nu_1\Delta t\sqrt{2\pi\eta\gamma\Delta t}}\right)}, \\
        &\approx \sqrt{\left(\frac{2}{\eta\gamma\Delta t}\right)\ln\left(\frac{1}{\nu_1\Delta t\sqrt{2\pi\eta\gamma\Delta t}}\right)}.
    \end{aligned}
\end{equation}
From Eq.~(\ref{appeq:approx_rates}), we have
\begin{equation}
    \nu_0 \approx \nu_1 \exp\left[-\sqrt{(2\eta\gamma\Delta t)\ln\left(\frac{1}{\nu_1\Delta t \sqrt{2\pi\eta\gamma\Delta t}}\right)}\right].
\end{equation}

\end{document}